\documentclass[12pt,a4paper,reqno]{amsart}
\usepackage{a4}
\usepackage{array}
\usepackage{eqnarray}
\usepackage{bbm}
\usepackage{mathrsfs}
\usepackage{cite}

\def\ot{\otimes}
\def\op{\oplus}
\def\t{\tilde}
\def\h{\hat}
\def\diag{\mathrm{diag}\,}
\def\id{\mathrm{i}\!\;\!\mathrm{d}}
\def\tr{\mathrm{tr}\!\:}
\def\iu{\mathrm{i}}
\def\bfd{\mathbf{d}}
\def\sfD{\mathsf{D}}
\def\bsj{\boldsymbol{\psi}}
\def\bsu{\boldsymbol{u}}
\def\bsd{\boldsymbol{d}}
\def\bfA{\mathbf{A}}
\def\one{\mathbbm{1}}
\def\CX{{C^\infty (X)}}

\def\C{\mathbbm{C}}
\def\L#1{\mathscr{L}_{#1}}
\def\M{\mathcal{M}}
\def\R{\mathbbm{R}}

\def\su#1{\mathrm{su(#1)}}
\def\u#1{\mathrm{u(#1)}}
\def\yn{\yesnumber}
\def\npb{\nopagebreak}
\def\be{\begin{equationarray*}}
\def\ee{\end{equationarray*}}
\def\hs{\hspace}
\def\vs{\vspace}
\def\tsum{\textstyle \sum}

\newcommand{\ba}[2][\!\!\!]{\left( #1 \begin{array}{#2}}
\newcommand{\ea}[1][\!\!\!]{\end{array} #1 \right)} 
\newcommand{\bb}[2][\!]{\left[ #1 \begin{array}{#2}}
\newcommand{\eb}[1][\!]{\end{array} #1 \right]} 
\newcolumntype{Z}[1]{>{$\hfil}m{#1mm}<{$\hfil}>{$\hfil}m{#1mm}<{$\hfil}}
\newcolumntype{E}[1]{>{$\hfil}m{#1mm}<{$\hfil}}

\newcommand{\W}[2][g]{\Omega^{#2} \mathfrak{#1}}
\newcommand{\p}[2][a]{\hat{\pi}(\Omega^{#2}\mathfrak{#1})}
\newcommand{\cJ}[2][g]{\mathbbm{J}^{#2}\!\mathfrak{#1}}
\newcommand{\cj}[2][a]{\mathbbm{j}^{#2}\!\mathfrak{#1}}
\newcommand{\bbr}[2][a]{\mathbbm{r}^{#2}\!\mathfrak{#1}}
\newcommand{\jj}[2][a]{\hat{\pi}(\mathcal{J}^{#2}{\mathfrak #1})}
\newcommand{\pf}[1][a]{\hat{\pi}({\mathfrak #1})}
\newcommand{\f}[1][g]{\ensuremath{\mathfrak{#1}}}
\newcommand{\mat}[2][\C]{{\mathrm{M}}_{#2}{#1}}
\newcommand{\rf}[1][**]{(\ref{#1})}
\newcommand{\g}[1][5]{\gamma^{#1}}

\makeatletter
\def\@setaddresses{\par
  \nobreak \begingroup
\footnotesize
  \def\author##1{\nobreak\addvspace\bigskipamount}%
  \def\\{\unskip, \ignorespaces}%
  \interlinepenalty\@M
  \def\address##1##2{\begingroup \centering \renewcommand{\arraystretch}{1}
    \par\addvspace\bigskipamount\indent
    \@ifnotempty{##1}{(\ignorespaces##1\unskip) }%
    {\scshape\ignorespaces \begin{tabular}[t]{c}##2%
			   \end{tabular}}\par\endgroup}%
  \def\curraddr##1##2{\begingroup
    \@ifnotempty{##2}{\nobreak\indent{\itshape Current address}%
      \@ifnotempty{##1}{, \ignorespaces##1\unskip}\/:\space
      ##2\par\endgroup}}%
  \def\email##1##2{\begingroup\centering
    \@ifnotempty{##2}{\nobreak\indent{\itshape E-mail}%
      \@ifnotempty{##1}{, \ignorespaces##1\unskip}\/:\space
      \ttfamily##2\par\endgroup}}%
  \addresses
  \endgroup
}
\def\@maketitle{%
  \normalfont\normalsize
  \let\@makefnmark\relax  \let\@thefnmark\relax
  \ifx\@empty\@date\else \@footnotetext{\@setdate}\fi
  \ifx\@empty\@subjclass\else \@footnotetext{\@setsubjclass}\fi
  \ifx\@empty\@keywords\else \@footnotetext{\@setkeywords}\fi
  \ifx\@empty\thankses\else \@footnotetext{%
    \def\par{\let\par\@par}\@setthanks}\fi
  \@mkboth{\@nx\shortauthors}{\@nx\shorttitle}%
  \global\topskip42\p@\relax 
  \@settitle
  \ifx\@empty\authors \else \@setauthors \fi
  \ifx\@empty\addresses \else \@setaddresses \fi
  \ifx\@empty\@dedicatory
  \else
    \baselineskip18\p@
    \vtop{\centering{\footnotesize\itshape\@dedicatory\@@par}%
      \global\dimen@i\prevdepth}\prevdepth\dimen@i
  \fi
  \@setabstract
  \normalsize
  \if@titlepage
    \newpage
  \else
    \dimen@34\p@ \advance\dimen@-\baselineskip
    \vskip\dimen@\relax
  \fi
} 
\let\enddoc@text\empty
\makeatother

\renewcommand{\arraystretch}{1.2}

\begin{document}

\title	 { The Standard Model within Non--associative Geometry}
\author  { Raimar Wulkenhaar }
\address { Institut f\"ur Theoretische Physik, Universit\"at Leipzig \\
	   Augustusplatz 10/11, D--04109 Leipzig, Germany} 
\email	 { wulkhaar@tph100.physik.uni-leipzig.de}
\date	 { \today}

\begin{abstract}
We present the construction of the standard model within the framework of 
non--associative geometry. For the simplest scalar product we get the 
tree--level predictions $m_W=\frac{1}{2} m_t\,,$ $m_H=\frac{3}{2} m_t$ and 
$\sin^2 \theta_W= \frac{3}{8}.$ These relations differ slightly from 
predictions derived in non--commutative geometry.
\end{abstract}

\maketitle 

\section{Introduction}

One of the most important applications of non--commutative geometry \cite{ac} 
to physics is a unified description of the standard model. The most elegant 
version rests upon a K--cycle \cite{cl, ac} with real structure \cite{acr}, see 
\cite{iks, mgv} for details and \cite{ks4, ks} for an older version. 
There also exist numerous other formulations within non--commutative geometry 
(NCG), see for instance \cite{ces, cev}. The author of this paper has proposed 
in \cite{rw2} a modification of non--commutative geometry. In that approach one 
uses unitary Lie algebras instead of unital associative $*$--algebras. Lie 
algebras are non--associative algebras~-- this is the motivation for the 
working title ``non--associative geometry''. The only realistic physical 
model that one can construct within the most elegant NCG--prescription is the 
standard model \cite{lmms}. The advantage of non--associative geometry is that 
a larger class of physical models can be constructed from the same amount of 
structures as in the most elegant NCG--formulation. That class includes the 
standard model, as we show in this paper. 

We give in Section~\ref{nag} a recipe how to construct classical gauge field 
theories within non--associative geometry. The arguments why this recipe 
works can be found in \cite{rw2}. Section~\ref{tsm} contains the construction 
of the standard model. We derive the geometric structures and write down the 
bosonic action for the simplest scalar product. The fermionic action will not 
be displayed, because it is identical with the classical formulation.

\section{The Recipe of Non--associative Geometry}
\label{nag}

The basic object in non--associative geometry is an L--cycle 
$(\f[g],h,D,\pi,\Gamma)\,,$ which consists of a $*$--representation $\pi$ of a 
unitary Lie algebra $\f[g]$ in bounded operators on a Hilbert space $h\,,$ 
together with a selfadjoint operator $D$ on $h$ with compact resolvent and a 
selfadjoint operator $\Gamma$ on $h\,,$ $\Gamma^2=\id_h\,,$ which commutes with 
$\pi(\f[a])$ and anticommutes with $D\,.$ The operator $D$ may be unbounded, 
but such that $[D,\pi(\f[g])]$ is bounded. L--cycles are naturally related to 
physical models if the following input data are given:
\begin{enumerate}
\vs{-\topsep} 
\item
The (Lie) group of local gauge transformations $\mathscr{G}\,.$  \label{sybr1}
\vs{-\itemsep} \vs{-\parsep}
\item 
Chiral fermions $\bsj$ transforming under a representation $\t{\pi}$ 
of $\mathscr{G}\,.$ 
\vs{-\itemsep} \vs{-\parsep}
\item
The fermionic mass matrix $\widetilde{\M}\,,$ i.e.\ fermion masses plus 
generalized Kobayashi--Maskawa matrices. 
\label{sybr3}
\vs{-\itemsep} \vs{-\parsep}
\item
Possibly the symmetry breaking pattern of $\mathscr{G}\,.$  \label{sybr4}
\end{enumerate}
\vs{-\topsep}
Take $\f[g]=\CX \ot \f[a]$ as the Lie algebra of $\mathscr{G}\,,$ where $\f[a]$ 
is a matrix Lie algebra and $\CX$ the algebra of smooth functions on the 
(compact Euclidian) space--time manifold $X\,.$ Take $h=L^2(X,S) \ot \C^F$ as 
the completion of the Euclidian fermions, where $L^2(X,S)$ is the Hilbert 
space of square integrable bispinors. Take $\pi=1 \ot \h{\pi}$ as the 
differential $\t{\pi}_*\,,$ where $\h{\pi}$ is a representation of $\f[a]$ in 
$\mat{F}\,.$ Put $D=\sfD \ot \one_F + \g \ot \M\,,$ where $\sfD$ is the Dirac 
operator on $X$ and $\M \in \mat{F}$ such that $\g \ot \M$ coincides with 
$\widetilde{\M}$ on chiral fermions. The chirality properties of the 
fermions are encoded in $\Gamma=\g \ot \h{\Gamma}.$ 

The recipe towards the (classical) gauge field theory associated to the 
L--cycle is the following: Let $\W[a]{1}$ be the space of formal commutators
\[
\omega^1=\tsum_{\alpha,z \geq 0} [a^z_\alpha,\dots [a^1_\alpha,d a^0_\alpha] 
\dots ]~,~~ a^i_\alpha \in \f[a]~.
\]
Apply linear mappings $\h{\pi}: \W[a]{1} \to \mat{F}$ and $\h{\sigma}: 
\W[a]{1} \to \mat{F}$ defined by
\begin{align}
\h{\pi}(\omega^1) &:= \tsum_{\alpha,z \geq 0} [\h{\pi}(a^z_\alpha),\dots 
[\h{\pi}(a^1_\alpha), [-\iu \M, \h{\pi}(a^0_\alpha)]] \dots ] ~, \\
\h{\sigma}(\omega^1) &:= \tsum_{\alpha,z \geq 0} [\h{\pi}(a^z_\alpha),\dots 
[\h{\pi}(a^1_\alpha), [\M^2, \h{\pi}(a^0_\alpha)]] \dots ] ~.
\label{hsig}
\end{align}
Define $\W[a]{n} \supset \omega^n=\tsum_\alpha [\omega^1_{n,\alpha}, 
[ \omega^1_{n-1,\alpha}, \dots [\omega^1_{2,\alpha}, \omega^1_{1,\alpha}] 
\dots ]]\,,$ where $\omega^1_{i,\alpha} \in \W[a]{1}\,.$ 
Extend $\h{\pi}$ and $\h{\sigma}$ recursively to $\W[a]{n}$ by 
\begin{align*}
\h{\pi}([\omega^1,\omega^k]) &:= \h{\pi}(\omega^1) \h{\pi}(\omega^k) 
- (-1)^k \h{\pi}(\omega^k) \h{\pi}(\omega^1) ~, \\
\h{\sigma}([\omega^1,\omega^k]) &:= \h{\sigma}(\omega^1) \h{\pi}(\omega^k) 
- \h{\pi}(\omega^k) \h{\sigma}(\omega^1) 
- \h{\pi}(\omega^1) \h{\sigma}(\omega^k) - (-1)^k \h{\sigma}(\omega^k) 
\h{\pi}(\omega^1) ~.
\end{align*}
Define for $n \geq 2$ 
\begin{equation}
\jj{n} :=\{~ \h{\sigma}(\omega^{n-1})~,~~ \omega^{n-1} \in \W[a]{n-1} \cap \ker 
\h{\pi}~\} ~. \label{js}
\end{equation}
Define spaces $\bbr{0} \subset \mat{F}$ and $\bbr{1} \subset \mat{F}$ 
elementwise by 
\begin{align}
\bbr{0} &= -(\bbr{0})^* = \h{\Gamma} (\bbr{0}) \h{\Gamma} ~, &
\bbr{1} &= -(\bbr{1})^* = -\h{\Gamma} (\bbr{1}) \h{\Gamma} ~, \notag \\{}
[\bbr{0} , \pf] &\subset \pf~, & [\bbr{0} , \p{1}] &\subset \p{1} ~, 
\label{rlh}\\
\{\bbr{0} , \pf\} &\subset \{ \pf, \pf\} + \p{2}~, &
\{\bbr{0} , \p{1}\} &\subset \{ \pf, \p{1}\} + \p{3}~, \hs*{-3em} \notag\\{}
[\bbr{1} ,\pf] &\subset \p{1}~, &
\{\bbr{1} ,\p{1}\} &\subset \p{2} + \{ \pf, \pf\} ~.  \notag
\end{align}
Define spaces $\cj{0},\cj{1},\cj{2} \subset \mat{F}$ elementwise by 
\be{rclrcl}
\multicolumn{6}{l}{
\cj{0}=-(\cj{0})^* = \h{\Gamma} (\cj{0}) \h{\Gamma}\;, \quad
\cj{1}=-(\cj{0})^* =-\h{\Gamma} (\cj{0}) \h{\Gamma}\;, \quad
\cj{2}= (\cj{0})^* = \h{\Gamma} (\cj{0}) \h{\Gamma}\;, }
\npb \\{}
[\cj{0}, \pf]	 &=& 0~, & \{\cj{0}, \pf\}   &\subset& \jj{2} + \{\pf, \pf\} ~, 
\npb \\{}
[\cj{0}, \p{1}]  &=& 0~, & \{\cj{0}, \p{1}\} &\subset& \jj{3} + \{\p{1},\pf\}~, 
\\{}
[\cj{1}, \pf]	 &=& 0~, & \{\cj{1}, \p{1}\} &\subset& \jj{2} + \{\pf, \pf\} ~, 
\npb \\{}
\{\cj{1}, \pf\}  & \subset& \multicolumn{4}{l}{\jj{3} + \{\p{1},\pf\} ~, } 
\label{cj} \yn \npb \\{} 
[\cj{1}, \p{1} ] & \subset& \multicolumn{4}{l}{ \jj{4} + \{\p{2} +\{\pf,\pf\} , 
		    \pf\} + } \npb \\
&+& \multicolumn{4}{l}{[\p{1}, \{ \pf,\pf\} ] 
+ \h{\sigma}(\h{\pi}^{-1}(\{\pf,\p{1}\} \cap \p{3})) ~, } \npb \\{} 
[\cj{2}, \pf ] & \subset& \jj{2} + \{ \pf, \pf\} ~, \hs*{1.5em} & 
[\cj{2}, \p{1} ] & \subset& \jj{3} + \{\p{1},\pf\} ~.  
\ee
The connection form $\rho$ has the structure
\begin{equation}
\rho = \tsum_\alpha (c^1_\alpha \ot m^0_\alpha + c^0_\alpha \g \ot 
m^1_\alpha) ~,~~ c^1_\alpha \in \Lambda^1~,~~ c^0_\alpha \in \Lambda^0~,~~ 
m^0_\alpha \in \bbr{0}~,~~ m^1_\alpha \in \bbr{1}~,~~ \label{rho} 
\end{equation}
where $\Lambda^k$ is the space of differential $k$--forms represented by gamma 
matrices. The curvature $\theta$ is computed from the connection form $\rho$ by 
\begin{equation}
\begin{split}
\label{th}
\theta &= \bfd \rho + \rho^2 - \iu \{ \g \ot \M, \rho\} + \h{\sigma}(\rho) \g
+ \cJ{2}~, \\
\cJ{2} &= (\Lambda^2 \ot \cj{0}) \op (\Lambda^1 \g \ot \cj{1}) \op 
(\Lambda^0 \ot \cj{2})~, 
\end{split}
\end{equation}
where $\bfd$ is the exterior differential and $\h{\sigma}$ the extension to 
elements of the form \rf[rho]. Select the representative $\f[e](\theta)$ 
orthogonal to $\cJ{2}\,,$ i.e.\ find $\mathrm{j} \in \cJ{2}$ such that
\begin{align*}
\f[e](\theta) &= \bfd \rho + \rho^2 - \iu \{ \g \ot \M, \rho\} 
+ \h{\sigma}(\rho) \g+ \mathrm{j}~,~~ & 
\int_X \!\! dx \;\tr(\f[e](\theta) j^2) &= 0~~\forall j^2 \in \cJ{2}~.
\end{align*}
The trace includes the trace in $\mat{F}$ and over gamma matrices. Compute the 
bosonic and fermionic actions 
\begin{align}
S_B &= \int_X \!\! dx \; \frac{1}{g_0^2\,F} \, \tr ( \f[e](\theta)^2 )~, &
S_F &= \int_X \!\! dx \; \bsj^* (D+\iu \rho) \bsj ~, \label{sb} 
\end{align}
where $g_0$ is a coupling constant and $\bsj \in h\,.$ Finally, 
perform a Wick rotation to Minkowski space. 

\section{The Construction}
\label{tsm}

Our constructions requires that the mass matrices of all fermions of the same 
type, including the neutrinos, are different from zero and non--degenerated. In 
particular, the Kobayashi--Maskawa matrix in both the quark and the lepton 
sector must be non--trivial. This is necessary to avoid certain degeneracy 
effects. The matrix Lie algebra of the standard model is 
\[
\f[a]=\su3 \op \su2 \op \u1~.~~
\]
The Hilbert space is $\C^{48}\,,$ because we need right neutrinos. We label 
elements of $\C^{48}$ in a suggestive way by the fermions 
of the first generation:
\[
(\bsu_L\,,\,\bsd_L\,,\,\bsu_R\,,\,\bsd_R\,,\,
{\nu}_L\,,\,e_L\,,\,{\nu}_R\,,\,e_R)^T \in \C^{48}~,~~\yn
\]
where $\bsu_L,\bsd_L,\bsu_R,\bsd_R \in \C^3 \ot \C^3$ and ${\nu}_L,e_L,{\nu}_R, 
e_R \in \C^3\,.$ The representation $\h{\pi}$ of $\f[a]$ on $\C^{48}$ is 
\be{l}
\h{\pi}((a_1,a_2,a_3))= \yn \label{sm0} \npb \\ \hs*{2em}
\iu f_0 \, \diag(\tfrac{1}{3} \one_3 \ot \one_3\,,\,
\tfrac{1}{3} \one_3 \ot \one_3\,,\, \tfrac{4}{3} \one_3 \ot \one_3\,,\,
-\tfrac{2}{3} \one_3 \ot \one_3\,,\, - \one_3\,,\, - \one_3\,,\, 0_3 \,,\,
-2 \one_3 ) +  \npb \\
\mbox{\small{$
\ba{c|c}
\begin{array}{cc|c|c} 
(a_3+\iu f_3 \one_3) \ot \one_3\,;~~ & \iu(f_1-\iu f_2) \one_3 \ot \one_3 & 
0 & 0 \\ 
\iu(f_1+\iu f_2) \one_3 \ot \one_3\,;~~ & (a_3-\iu f_3 \one_3) \ot \one_3 & 
0 & 0 \\ \hline
0 & 0 & \,a_3 \ot \one_3\, & 0 \\ \hline 
0 & 0 & 0 & \,a_3 \ot \one_3\, \end{array} & 
\mathrm{O} \\ \hline 
\mathrm{O} & \begin{array}{cc|E{6}|E{6}}
\multicolumn{1}{l}{\iu f_3 \ot \one_3 \,;} & \hs*{-0.8em} \iu(f_1-\iu f_2) \ot 
\one_3 & 0 & 0 \\ \iu(f_1+\iu f_2) \ot \one_3\, ; \hs*{-0.5em} & 
\multicolumn{1}{r}{-\iu f_3 \ot \one_3 } \vline & 0 & 0 \\ \hline 
0 & 0 & 0_3 & 0 \\ \hline
0 & 0 & 0 & 0_3 \end{array} \ea $}}\!.
\ee
Here, the matrix $a_3 \in \su3 \subset \mat{3}$ is written down in the standard 
matrix representation, ${\displaystyle a_2 = \binom{ \iu f_3\, ; 
\qquad \iu(f_1-\iu f_2) }{ \iu(f_1+\iu f_2)\, ; \quad -\iu f_3 }} \in \su2 
\,,$ for $f_1,f_2,f_3\in \R\,,$ and $a_1=\iu f_0 \in \u1 \equiv \iu \R\,.$
The generalized Dirac operator is
\[
\M=\mbox{\small{$
\ba{c|c}
\begin{array}{Z{15}|E{15}|E{15}}   
0 & 0 & \one_3 \ot M_u & 0 \\ 
0 & 0 & 0 & \one_3 \ot M_d \\ \hline
\one_3 \ot M_u^* & 0 & 0 & 0 \\ \hline
0 & \one_3 \ot M_d^* & 0 & 0 \end{array} & \mathrm{O} \\ \hline 
\mathrm{O} & \begin{array}{Z{10}|E{10}|E{10}}
0 & 0 & M_{\nu} & 0 \\
0 & 0 & 0 & M_e \\ \hline
M_{\nu}^* & 0 & 0 & 0 \\ \hline
0 & M_e^* & 0 & 0 \end{array}
\ea $}},~~ \yn
\]
where $M_u,M_d,M_{\nu},M_e \in \mat{3}$ are the mass matrices of the fermions. 
It is easy to see that for $a^i_\alpha=(a^i_{3,\alpha},a^i_{2,\alpha}, 
a^i_{3,\alpha}) \in \f[a]$ one has
\enlargethispage{10mm}
\be{l}
\tau^1:=\tsum_{\alpha,z \geq 0} [\h{\pi}(a^z_\alpha),\dots 
[\h{\pi}(a^1_\alpha), [-\iu\M,\h{\pi}(a^0_\alpha)]] \dots ] = 
\yn \label{sm1a} \npb \\
\iu \! \mbox{\small{$ \ba{c|c} 
\begin{array}{Z{19.5}|E{19}|E{18}}	 
0 & 0 &  \bar{b}_2 \one_3 \ot M_u & b_1 \one_3 \ot M_d \\ 
0 & 0 & -\bar{b}_1 \one_3 \ot M_u & b_2 \one_3 \ot M_d \\ \hline
b_2 \one_3 \ot M_u^*\,; & -b_1 \one_3 \ot M_u^* & 0 & 0 \\ \hline
\bar{b}_1 \one_3 \ot M_d^*\,; & \bar{b}_2 \one_3 \ot M_d^* & 0 & 0 \end{array} 
& \mathrm{O} \\ \hline
\mathrm{O} & \begin{array}{Z{15.5}|E{15}|E{13}}
0 & 0 & \bar{b}_2 \ot M_{\nu} & b_1 \ot M_e \\
0 & 0 & -\bar{b}_1 \ot M_{\nu} & b_2 \ot M_e \\ \hline
b_2 \ot M_{\nu}^*\,; & -b_1 \ot M_{\nu}^* & 0 & 0 \\ \hline
\bar{b}_1 \ot M_e^*\,; & \bar{b}_2 \ot M_e^* & 0 & 0 \end{array}
\ea $}}\!,
\\
\binom{b_1}{b_2} = \sum_{\alpha,z \geq 0} a^z_{2,\alpha} a^z_{1,\alpha} \cdots 
a^1_{2,\alpha} a^1_{1,\alpha} a^0_{2,\alpha} a^0_{1,\alpha} \binom{0}{1} 
\in \C^2~.~~  \yn \label{sm1b}
\ee
The matrix \rf[sm1a] is the general form of an element of $\p{1}\,.$ 
The grading operator is 
\[
\h{\Gamma}=\diag ( -\one_{3} \ot \one_3\,,\, -\one_{3} \ot \one_3\,,\, 
\one_{3} \ot \one_3\,,\, \one_{3} \ot \one_3\,,\, - \one_3\,,\, - \one_3\,,\, 
\one_3\,,\, \one_3 )~.~~  \label{gr2} \yn
\]
One has $\h{\Gamma}^2 = \one_{48}\,,$ $[\h{\Gamma},\pf]=0\,,$ 
$\{\h{\Gamma},\M\}=0$ and $\{\h{\Gamma},\p{1}\}=0\,.$ Let 
\be{l}
\binom{\iu f_3\,; \qquad \iu(f_1-\iu f_2) }{ \iu(f_1+\iu f_2)\,; 
\quad -\iu f_3} := \binom{\iu (|b_2|^2 - |b_1|^2)\,; \qquad 
- 2 \iu b_1 \bar{b}_2 } {-2 \iu \bar{b}_1 b_2\,; \quad 
- \iu (|b_2|^2 - |b_1|^2) } \in \su2~,  \\
M_{ud}=M_u M_u^* - M_d M_d^*~,~~M_{\nu e}=M_{\nu} M_{\nu}^* - M_e M_e^*~,~~ \\
M_{\{ud\}}=M_u M_u^* + M_d M_d^*~,~~M_{\{\nu e\}}=M_{\nu} M_{\nu}^* 
+ M_e M_e^*  ~.
\ee
Then we have 
\be{l}
\{\tau^1,\tau^1\}= \npb \\ 
\iu \mbox{\small{$\ba{c|c} \,
\begin{array}{cc|E{5.5}|E{5.5}} 
\multicolumn{1}{l}{\,\iu f_3 \one_3 \!\ot\! M_{ud}\,; } & 
\hs*{-0.55em} \iu(f_1-\iu f_2) \one_3 \!\ot\! M_{ud} \, & 0 & 0 \\ 
\,\iu(f_1+\iu f_2) \one_3 \!\ot\! M_{ud}\,; \hs*{-0.55em}  & 
\multicolumn{1}{r}{-\iu f_3 \one_3 \!\ot\! M_{ud}\,} \vline & 0 & 0 \\ \hline
0 & 0 & 0_9 & 0 \\ \hline 
0 & 0 & 0 & 0_9 \end{array} & 
\mathrm{O} \\ \hline 
\mathrm{O} & \begin{array}{cc|E{5.5}|E{5.5}}
\multicolumn{1}{l}{\,\iu f_3 \!\ot\! M_{\nu e}\,;} & \hs*{-0.55em} 
\iu(f_1-\iu f_2) \!\ot\! M_{\nu e} \,& 0 & 0 \\
\, \iu(f_1+\iu f_2) \!\ot\! M_{\nu e}\,; \hs*{-0.55em} & \multicolumn{1}{r}{ 
-\iu f_3 \!\ot\! M_{\nu e}\, } \vline & 0 & 0 \\ \hline 
0 & 0 & 0_3 & 0 \\ \hline
0 & 0 & 0 & 0_3 \end{array} \, \ea $}} \hs*{2em} \yn \label{sm2a}
\\
-(\|b_1\|^2 \!+\! \|b_2\|^2)\, \diag \big( 
\one_3 \ot M_{\{ud\}} \,,\, \one_3 \ot M_{\{ud\}} \,,\, \one_3 \ot 2 M_u^* M_u 
\,, \, \one_3 \ot 2 M_d^* M_d \,, ~~\\ 
\multicolumn{1}{r}{ M_{\{\nu e\}} \,,\, M_{\{\nu e\}} \,,\, 
2 M_{\nu}^* M_{\nu} \,,\, 2 M_e^* M_e \big)\;.\hs*{2em} } \yn \label{sm2b}
\ee
Next, for $\tau^1=\h{\pi}(\omega^1)$ given by \rf[sm1a] we obtain with 
\rf[hsig] 
\be{l}
\h{\sigma}(\omega^1) = \tsum_{\alpha,z \geq 0} [\h{\pi}(a^z_\alpha),\dots 
[\h{\pi}(a^1_\alpha), [\M^2,\h{\pi}(a^0_\alpha)]] \dots ] = 
\yn \label{sm2} \npb \\
\iu \mbox{\small{$\ba{c|c} \,
\begin{array}{cc|E{5.5}|E{5.5}} 
\multicolumn{1}{l}{\,\iu f_3 \one_3 \!\ot\! M_{ud}\,; \!\!} & 
\!\! \iu(f_1-\iu f_2) \one_3 \!\ot\! M_{ud} \, & 0 & 0 \\ 
\, \iu(f_1+\iu f_2) \one_3 \!\ot\! M_{ud}\,; \!\! & 
\multicolumn{1}{r}{\!\! -\iu f_3 \one_3 \!\ot\! M_{ud} \,} \vline & 0 & 0 
\\ \hline
0 & 0 & 0_9 & 0 \\ \hline 
0 & 0 & 0 & 0_9 \end{array} & 
\mathrm{O} \\ \hline 
\mathrm{O} & \begin{array}{cc|E{5.5}|E{5.5}}
\multicolumn{1}{l}{\, \iu f_3 \!\ot\! M_{\nu e}\,;} & \iu(f_1-\iu f_2) 
\!\ot\! M_{\nu e} \, & 0 & 0 \\
\, \iu(f_1+\iu f_2) \!\ot\! M_{\nu e}\,; & \multicolumn{1}{r}{ 
-\iu f_3 \!\ot\! M_{\nu e} \, } \vline & 0 & 0 \\ \hline 
0 & 0 & 0_3 & 0 \\ \hline
0 & 0 & 0 & 0_3 \end{array} \, \ea $}},~~ \\
\binom{\iu f_3\,; \qquad \iu(f_1-\iu f_2) }{ \iu(f_1+\iu f_2)\,; \quad 
-\iu f_3} :=\!\! \sum_{\alpha,z \geq 0} \!\! \big[ a^z_{2,\alpha} ,\dots \big[ 
a^1_{2,\alpha}, \big[ a^0_{2,\alpha}, \binom{\tfrac{\iu}{2} \qquad 0 }{ 0 \quad 
- \tfrac{\iu}{2} } \big] \big] \dots \big] \in \su2~.~~ 
\ee
Choosing 
\be{l}
\omega^1_0=da_2^0+[a_2^1,[a_2^1,da_2^0] ~,~~
a_2^0= \binom{0 \quad \iu }{ \iu \quad 0} \in \su2~,~~
a_2^1= \binom{\iu \qquad 0 }{ 0 \quad -\iu } \in \su2~,
\ee
we have $\omega^1_0 \in \ker \h{\pi}$ due to $(a_2^0 + a_2^1 a_2^1 a_2^0) 
\binom{0}{1} = \binom{0}{0}\,,$ see \rf[sm1b]. On the other hand, 
$\h{\sigma}(\omega^1) \neq 0$ is the matrix \rf[sm2], with 
$f_1=0,f_2=-3,f_3=0\,.$ Obviously, each matrix of the form \rf[sm2] can be 
represented as $\h{\sigma}(\omega^1)\,,$ for $\omega^1=\tsum_{\alpha,z \geq 0} 
[a^z_\alpha,\dots [a^0_\alpha, \omega^1_0] \dots ] \in \ker \h{\pi}\,.$ 
Therefore, each element of $\jj{2}$ is precisely of the form 
\rf[sm2], see \rf[js]: 
\be{c}
\h{\sigma}(\W[a]{1}) \equiv \jj{2}~.~~	\label{wgj} \yn
\ee
Comparing the results \rf[wgj] and \rf[sm2] with \rf[sm2a] and \rf[sm2b] we get 
\be{rl}
\{\tau^1 \! ,\tau^1\} = -(\|b_1\|^2 \!+\! \|b_2\|^2)\, \diag & \big( 
\one_3 \ot M_{\{ud\}} \,, \one_3 \ot M_{\{ud\}} \,, \one_3 \ot 2 M_u^* M_u 
\,, \one_3 \ot 2 M_d^* M_d \,, \\ 
& M_{\{\nu e\}} \,,\, M_{\{\nu e\}} \,,\, 
2 M_{\nu}^* M_{\nu} \,,\, 2 M_e^* M_e \big) \mod \jj{2} \;. ~~\yn 
\label{mj2}
\ee
It is clear that \rf[mj2] is orthogonal to $\jj{2}\,.$ 

Next, we need the structure of the space $\{\pf,\pf\}\,.$ A simple 
calculation yields for elements of $\{\pf,\pf\}$ the form 
\be{l}
\{\pf, \pf\} \ni \diag( A_q + \Delta_q\,,\, A_\ell + \Delta_\ell)~,~~ \\
A_q= \sum_\alpha \iu \mbox{\small{$\ba{lr|c|c} 
\bb{c} (\tfrac{1}{3} \h{\lambda}^0_\alpha \!+\! \lambda^3_\alpha 
\!+\! \lambda^0_\alpha) a_{3,\alpha} \\ 
+ \tfrac{1}{3} \iu \h{\lambda}^3_\alpha \one_3 \eb{} \hs*{-0.6em} & 
\bb{c} (\lambda^1_\alpha -\iu \lambda^2_\alpha) a_{3,\alpha} + \\
\tfrac{1}{3} \iu (\h{\lambda}^1_\alpha - \iu \h{\lambda}^2_\alpha) \one_3 
\eb{} & 0 & 0 
\\ 
\bb{c} (\lambda^1_\alpha +\iu \lambda^2_\alpha) a_{3,\alpha} + \\ 
\tfrac{1}{3} \iu (\h{\lambda}^1_\alpha + \iu \h{\lambda}^2_\alpha) 
\one_3 \eb{} & \hs*{-0.6em} \bb{c} (\tfrac{1}{3} \h{\lambda}^0_\alpha 
-\lambda^3_\alpha \!+\! \lambda^0_\alpha) a_{3,\alpha} \\ 
- \tfrac{1}{3} \iu \h{\lambda}^3_\alpha \one_3 \eb{} & 0 & 0 \\ \hline
\multicolumn{2}{c}{0 \hs*{7em} 0} \vline & 
(\lambda^0_\alpha \!+\! \tfrac{4}{3} \h{\lambda}^0_\alpha) a_{3,\alpha} & 0 \\ 
\hline
\multicolumn{2}{c}{0 \hs*{7em} 0} \vline & 
0 & (\lambda^0_\alpha - \tfrac{2}{3} \h{\lambda}^0_\alpha) a_{3,\alpha} \;
\ea \ot \one_3 \;,$}} 
\\
A_\ell= \sum_\alpha \iu \mbox{\small{$\ba{Z{30}|E{15}|E{15}} 
- \iu \h{\lambda}^3_\alpha \one_3 \,; & 
- \iu (\h{\lambda}^1_\alpha - \iu \h{\lambda}^2_\alpha) \one_3 & 0 & 0 \\ 
- \iu (\h{\lambda}^1_\alpha + \iu \h{\lambda}^2_\alpha) \one_3 &
\iu \h{\lambda}^3_\alpha \one_3 & 0 & 0 \\ \hline
0 & 0 & 0_3 & 0 \\ \hline
0 & 0 & 0 & 0_3 
\ea ,$}} \yn \label{smaa}
\\
\Delta_q= \diag \big( (\lambda \!+\! \t{\lambda} 
\!+\! \tfrac{1}{9} \h{\lambda}) \one_3 , 
(\lambda \!+\! \t{\lambda} \!+\! \tfrac{1}{9} \h{\lambda}) \one_3 , 
(\lambda \!+\! \tfrac{16}{9} \h{\lambda}) \one_3 , 
(\lambda \!+\! \tfrac{4}{9} \h{\lambda}) \one_3 \big) \ot \one_3\;,
\\
\Delta_\ell= \diag \big( (\t{\lambda} + \h{\lambda}) \one_3 \,,\, 
(\t{\lambda} + \h{\lambda}) \one_3 \,,\, 0_3\,,\, 4 \h{\lambda} \one_3 \big)~,
\ee
where $a_{3,\alpha} \in \su3$ and $\lambda^0_\alpha,\lambda^1_\alpha, 
\lambda^2_\alpha,\lambda^3_\alpha,\h{\lambda}^0_\alpha,\h{\lambda}^1_\alpha, 
\h{\lambda}^2_\alpha,\h{\lambda}^3_\alpha, \lambda,\t{\lambda},\h{\lambda} 
\in \R\,.$ 

\enlargethispage{5mm}
In order to write down the structure of the connection form we must find the 
spaces $\bbr{0}$ and $\bbr{1}\,,$ see \rf[rho]. The evaluation of \rf[rlh] 
yields in the case of generic mass matrices $M_u,M_d,M_{\nu},M_e$ the simple 
result
\begin{align*}
\bbr{0} &=\pf~,~~ & \bbr{1} &= \p{1}~.
\end{align*}
For generic mass matrices, equations \rf[cj] have the solution 
$\cj{0}=0\,,$ $\cj{1}=0$ and
\begin{equation}
\begin{split}
\cj{2} &= \jj{2} \op \{\pf,\pf\} \op \diag(\R \one_{18}\,,\,\R \one_{18}\,,\,
\R \one_{6}\,,\,\R \one_{6}) \\
& \subset J_2 \op \diag(A_q\,,\, A_\ell) \op \diag(J_q\,,\,J_\ell)~,~~ \\
J_q &= \diag((\lambda_1+\tfrac{1}{9} \lambda_0) \one_3\,,\, 
(\lambda_1+\tfrac{1}{9} \lambda_0) \one_3\,,\, 
(\lambda_2+\tfrac{16}{9} \lambda_0) \one_3\,,\, 
(\lambda_2+\tfrac{4}{9} \lambda_0) \one_3) \ot \one_3~,~~ \npb \\
J_\ell &= \diag((\lambda_3+ \lambda_0) \one_3\,,\, (\lambda_3+ \lambda_0) 
\one_3\,,\, \lambda_4 \one_3\,,\, (\lambda_4+4 \lambda_0) \one_3) ~,~~ 
\end{split}
\end{equation}
for $J_2 \in \jj{2}$ and $\lambda_1,\lambda_2,\lambda_3,\lambda_4 \in \R\,.$ 

In order to write down the bosonic action it is necessary to select the 
representative $\f[e](\{\tau^1,\tau^1\})$ of $\{\tau^1,\tau^1\} + \cj{2}$ 
orthogonal to $\cj{2}\,.$ This problem is easy to solve. Let 
\begin{equation*}
\begin{split}
\t{M}_{\{ud\}} &:= M_u M_u^* + M_d M_d^* - \tfrac{1}{3} 
\tr(M_u M_u^* + M_d M_d^* ) \one_3~,~~\\
\t{M}_{\{\nu e\}} &:= M_{\nu} M_{\nu}^* + M_e M_e^* -\tfrac{1}{3} 
\tr (M_{\nu} M_{\nu}^* + M_e M_e^* ) \one_3~,~~ \\
\t{M}_{uu} &:= M_u^* M_u - \tfrac{1}{24} \tr(5 M_u M_u^* + 3 M_d M_d^* 
- M_{\nu} M_{\nu}^* + M_e M_e^* ) \one_3~,~~  \\
\t{M}_{dd} &:= M_d^* M_d - \tfrac{1}{24} \tr(3 M_u M_u^* + 5 M_d M_d^* 
+ M_{\nu} M_{\nu}^* - M_e M_e^* ) \one_3~,~~  \\
\t{M}_{\nu\nu} &:= M_{\nu}^* M_{\nu} - \tfrac{1}{24} \tr(-3 M_u M_u^* 
+ 3 M_d M_d^* + 7 M_{\nu} M_{\nu}^* + M_e M_e^* ) \one_3~,~~  \\
\t{M}_{ee} &:= M_e^* M_e - \tfrac{1}{24} \tr(3 M_u M_u^* 
- 3 M_d M_d^* + M_{\nu} M_{\nu}^* + 7 M_e M_e^* ) \one_3~.~~
\end{split}
\end{equation*}
Then, the canonical embedding $\f[e](\{\tau^1,\tau^1\})$ of $\{\tau^1,\tau^1\}$ 
into $\mat{48}$ is given by
\be{rl}
\f[e](\{\tau^1,\tau^1\}) = -(\|b_1\|^2 \!+\! \|b_2\|^2)\, \diag & 
\big( \one_3 \ot \t{M}_{\{ud\}} \,,\, \one_3 \ot \t{M}_{\{ud\}} \,,\, 
\one_3 \ot 2 \t{M}_{uu} \,, \, \one_3 \ot 2 \t{M}_{dd} \,, ~~\npb \\ 
& \t{M}_{\{\nu e\}} \,,\, \t{M}_{\{\nu e\}} \,,\, 
2 \t{M}_{\nu\nu} \,,\, 2 \t{M}_{ee} \big) \,. ~~\yn
\ee

Now we include the four dimensional Riemannian spin manifold $X$ and choose a 
selfadjoint local basis $\{\g[\mu]\}_{\mu=1,2,3,4}$ of $\Lambda^1\,.$ The 
connection form $\rho$ has due to \rf[rho], \rf[sm0] and \rf[sm1a] the 
structure
\be{l}
\rho= \binom{\rho_q \qquad 0 }{ 0 \qquad \rho_{\ell} }~,~~ \yn \label{rh}
\\
\rho_q=\!\! \mbox{\small{$
\ba{cc|c|c} 
\,(\bfA+\iu (\tfrac{1}{3} A^0 + A^3) \one_3) \ot \one_3\,; \hs*{-1em} & 
\multicolumn{1}{r}{\iu(A^1-\iu A^2) \one_3 \ot \one_3\,} \vline & 
-\iu \g \bar{\Phi}_2 \one_3 \ot M_u & -\iu \g \Phi_1 \one_3 \ot M_d \\ 
\multicolumn{1}{l}{ \,\iu(A^1+\iu A^2) \one_3 \ot \one_3 \,;} & 
\hs*{-1em} (\bfA+\iu (\tfrac{1}{3} A^0 - A^3) \one_3) \ot \one_3\, & 
\iu \g \bar{\Phi}_1 \one_3 \ot M_u & -\iu \g \Phi_2 \one_3 \ot M_d \\ \hline
-\iu \g \Phi_2 \one_3 \ot M_u^* & \iu \g \Phi_1 \one_3 \ot M_u^* & 
\,(\bfA + \tfrac{4}{3} \iu A^0 \one_3) \ot \one_3\, & 0 \\ \hline 
-\iu \g \bar{\Phi}_1 \one_3 \ot M_d^* & -\iu \g \bar{\Phi}_2 \one_3 \ot M_d^* & 
0 & \,(\bfA - \tfrac{2}{3} \iu A^0 \one_3) \ot \one_3 ~~ \ea $}}\! , 
\\
\rho_{\ell}= \mbox{\small{$ 
\ba{Z{30}|E{25}|E{25}}
\iu (-A^0 + A^3) \ot \one_3  & \iu(A^1-\iu A^2) \ot \one_3 & 
-\iu \g \bar{\Phi}_2 \ot M_{\nu} & -\iu \g \Phi_1 \ot M_e \\ 
\iu(A^1+\iu A^2) \ot \one_3 & \iu (-A^0 - A^3) \ot \one_3 & 
\iu \g \bar{\Phi}_1 \ot M_{\nu} & -\iu \g \Phi_2 \ot M_e \\ \hline 
-\iu \g \Phi_2 \ot M_{\nu}^* & \iu \g \Phi_1 \ot M_{\nu}^* & 0_3 & 0 \\ \hline
-\iu \g \bar{\Phi}_1 \ot M_e^* & -\iu \g \bar{\Phi}_2 \ot M_e^* & 0 & 
-2 \iu A^0 \ot \one_3
\ea $}},
\ee
where $\bfA \in \Lambda^1 \ot \su3\,,$ $\t{\bfA} := {\displaystyle  
\binom{ \iu A^3 \,; \qquad \iu(A^1-\iu A^2) }{ \iu(A^1+\iu A^2) \,; \quad 
-\iu A^3 }} \in \Lambda^1 \ot \su2\,,$ $A^0 \in \Lambda^1\,,$ 
$\Phi_1,\Phi_2 \in \Lambda^0 \ot \C\,.$ In formula \rf[th] for the curvature 
note that $\h{\sigma}(\omega^1)=0 \mod \jj{2}\,.$ Inserting \rf[rh] into 
\rf[th] we obtain for the bosonic action given in \rf[sb] 
\enlargethispage{5mm}
\be{rcl}
S_B &=& \tfrac{1}{48 g_0^2} \int_X \!\!dx \;\tr (\f[e](\theta)^2) = 
\int_X \!\!dx \;(\L2 + \L1 + \L0)~,~~ \yn \label{Lsm}
\\
\L2 &=& \tfrac{1}{4 g_0^2} \tr ((\bfd \bfA + \tfrac{1}{2} \{\bfA, \bfA\})^2) 
+ \tfrac{1}{4 g_0^2} \tr ((\bfd \t{\bfA} + \tfrac{1}{2} 
\{\t{\bfA}, \t{\bfA}\})^2) + \tfrac{5}{6 g_0^2} \tr ((\bfd A^0)^2) ~, \\
\L1 &=& \begin{array}[t]{rl} \tfrac{1}{24 g_0^2} \tr \big( &
|\bfd \Phi_1 + \iu (A^0 + A^3) \Phi_1 + \iu (A^1-\iu A^2) (\Phi_2 +1)|^2  + 
\npb \\
+ & |\bfd \Phi_2 + \iu (A^0 - A^3) (\Phi_2+1) + \iu (A^1+\iu A^2) \Phi_1 |^2 
\big) \times \end{array} \npb \\ 
\multicolumn{3}{r}{\times \tr( 3 M_u M_u^* + 3 M_d M_d^* + M_{\nu} M_{\nu}^* 
+ M_e M_e^* )~,~~ } 
\npb \\
\L0 &=& \tfrac{1}{192 g_0^2} (|\Phi_1|^2 + |\Phi_2+1|^2-1)^2 \tr(1) \times  
\npb \\
\multicolumn{3}{r}{\times \tr( 6 \t{M}_{\{ud\}}^2 + 12 \t{M}_{uu}^2 
+ 12 \t{M}_{dd}^2 + 2 \t{M}_{\{\nu e\}}^2 + 4 \t{M}_{\nu\nu}^2 
+ 4 \t{M}_{ee}^2 ) ~.~~}
\ee
We perform the reparameterizations 
\begin{equation}
\begin{split}
\label{pra}
\bfA &=\tsum_{a=1}^8 \tfrac{\iu g_0}{2} G^a_{\mu} \g[\mu] \ot \lambda^a~,~~ 
\t{\bfA} =\tsum_{a=1}^3 \tfrac{\iu g_0}{2} W^a_{\mu} \g[\mu] \ot \sigma^a~,~~ 
A^0 = \tfrac{\iu g_0}{2} \sqrt{\tfrac{3}{5}} W^0_{\mu} \g[\mu] ~, 
\\
\Phi_i &= g_0 \phi_i\, / \, \sqrt{\tr( M_u M_u^* + M_d M_d^* + \tfrac{1}{3} 
M_{\nu} M_{\nu}^* + \tfrac{1}{3} M_e M_e^* ) } ~,~~ i=1,2~, 
\end{split}
\end{equation}
where $\{\sigma^a\}$ are the Pauli matrices and $\{\lambda^a\}$ the Gell-Mann 
matrices. Using 
\[
\tr( (\g[\kappa] \wedge \g[\lambda])(\g[\mu] \wedge \g[\nu]))= 
4(\delta^{\lambda\mu}\delta^{\kappa \nu} - \delta^{\kappa \mu } 
\delta^{\lambda\nu}) \;,~ \tr( \g[\mu] \g[\nu])=4 \delta^{\mu\nu}\;,~ 
\tr(1)=4 
\]
and performing a Wick rotation to Minkowski space we obtain for \rf[Lsm]
precisely the bosonic action of the standard model, see \cite{mg}. Here, 
the Weinberg angle $\theta_W$ and the masses $m_W, m_Z$ and $m_H$ of the 
$W$, $Z$ and Higgs bosons are given by
\begin{equation}
\begin{split}
m_W &= \tfrac{1}{2} \sqrt{\tr( M_u M_u^* + M_d M_d^* + \tfrac{1}{3} 
M_{\nu} M_{\nu}^* + \tfrac{1}{3} M_e M_e^* )} = \tfrac{1}{2} m_t~,~~  \\
m_Z &= m_W / \cos \theta_W~,~~\sin^2 \theta_W= \tfrac{3}{8}~,~~ \\
m_H &=\sqrt{\dfrac{\tr( \t{M}_{\{ud\}}^2 + 2 \t{M}_{uu}^2 
+ 2 \t{M}_{dd}^2 + \tfrac{1}{3} \t{M}_{\{\nu e\}}^2 + \tfrac{2}{3} 
M_{\nu\nu}^2 + \tfrac{2}{3} M_{ee} )}{ \tr( M_u M_u^* 
+ M_d M_d^* + \tfrac{1}{3} M_{\nu} M_{\nu}^* + \tfrac{1}{3} M_e M_e^* )}} 
= \tfrac{3}{2} m_t~,~~
\end{split}
\end{equation}
where $m_t$ is the mass of the top quark. Here we have neglected the other 
fermion masses against $m_t\,.$ The analogous relations in 
non--commutative geometry read for the simplest scalar product \cite{ks}
\begin{align}
m_W &=\tfrac{1}{2} m_t~, & m_H &=\sqrt{\tfrac{69}{28}}\,m_t~, &
\sin^2 \theta_W &=\tfrac{12}{29}~.  
\end{align}
Inserting \rf[rh] and \rf[pra] into the fermionic action in \rf[sb] we arrive 
after a Wick rotation to Minkowski space and imposing the chirality condition 
$\Gamma h= h$ at the usual fermionic action of the standard model \cite{mg}. 

\enlargethispage{10mm}

\end{document}